\documentclass[prd,aps,nofootinbib,singlecolumn]{revtex4-2}
\usepackage{amsmath,amssymb,amsfonts,color,graphicx,graphics,latexsym,placeins,epsfig,multirow}
\usepackage[toc]{appendix}
\usepackage{array}
\newcolumntype{P}[1]{>{\centering\arraybackslash}p{#1}}
\usepackage{footmisc}
\usepackage{mathrsfs}
\usepackage[compatibility=false]{caption}
\usepackage{subcaption}
\usepackage{comment}
\usepackage{bbold}
\usepackage{enumitem}
\setdescription{leftmargin=12.5pt}
\usepackage{bbold}
\usepackage{hyperref}
\usepackage{varioref}
\usepackage{color}
\usepackage{dirtytalk}
\def\a{\alpha}

\def\6{\partial}

\begin{document}

\title{\Large \bf Formulation of Galilean relativistic Born-Infeld theory}
\author{Rabin Banerjee$^1$ \footnote{DAE Raja Ramanna fellow}}
\author{Soumya Bhattacharya$^1$}
\author{Bibhas Ranjan Majhi$^2$}
\affiliation{$^1$Department of Astrophysics and High Energy Physics, S.N. Bose National Center for Basic Sciences, Kolkata 700106, India}
\email{rabin@bose.res.in}
\email{soumya557@bose.res.in}
\affiliation{$^2$ Department of Physics, Indian Institute of Technology Guwahati, Guwahati
781039, Assam, India}
\email{bibhas.majhi@iitg.ac.in}


\vskip .2in
\begin{abstract}
     In this paper, we formulate, for the first time, in a systematic manner, Galilean relativistic Born-Infeld action in detail. Exploiting maps connecting Lorentz relativistic and Galilean relativistic vectors, we construct the two limits (electric and magnetic) of Galilean relativistic Born-Infeld action from usual relativistic Born-Infeld theory. An action formalism is thereby derived. From this action,  equations of motion are obtained either in the potential or field formulation. Galilean version of duality transformations involving the electric and magnetic fields are defined. They map the electric limit relations to the magnetic ones and vice-versa, exactly as happens for Galilean relativistic Maxwell theory. We also explicitly show the Galilean boost and gauge invariances of the theory in both limits. 
\end{abstract}
\maketitle
\section{Introduction}
The low energy effective description of a system is usually explained by the study of a non-relativistic (NR) field theory. Non-relativistic limit of classical field theories received considerable attention recently. It has found applications in various diverse branches of theoretical physics like holography \cite{Taylor},  non-relativistic diffeomorphism invariant theories \cite{andreev1, rb13, rb14, jensen, rb1, rb2}, 
condensed matter physics \cite{Son, Pal}, fluid dynamics \cite{Jain, rb3} and gravitation \cite{Morand, Read} to name a few. This formulation is tricky and altogether different from the relativistic case. The absolute nature of time in the non-relativistic limit leads to the lack of a single non-degenerate metric which poses some additional difficulties. 
It is  very interesting to understand  Galilean invariance and its role in field theories, especially in gauge theories \cite{Sengupta}.
The basic construction of Galilean electrodynamics was first given by Le Bellac and Levy-Leblond \cite{Leblond} back in 1970's. This was done in the field formulation. A similar field based analysis was done in \cite{Khanna} using embedding techniques. Further directions in this type of analysis were provided in \cite{Duval}. A systematic covariant construction of the action principle of Galilean relativistic Maxwell theory was given in \cite{Bhattacharya} and a covariant action formulation of Galilean relativistic Proca theory was provided in \cite{Bhattacharya1}. Bellac and Levy-Leblond \cite{Leblond} (and others as well) only consider contravariant components. This
becomes necessary (considering both covariant and contravariant components) here since we have given an action formulation. Our approach
is to deduce the non-relatvistic expressions from the relativistic action
using the scalings. In this sense our approach is general that can be
developed either in terms of potentials or (electric and magnetic) fields.
This is not possible in other approaches.

The principal motivation of the present paper is to investigate Galilean versions of non-linear theories of electromagnetism. Among various possibilities, the Born-Infeld (BI) theory \cite{BI} is of particular importance. 
This is because the inclusion of electron's self energy demands non-linearity in the electromagnetic theory. In that case the effective action includes non-linear terms apart from the standard Maxwell term. Such an action is well known as Born–Infeld (BI) action \cite{BI} and gave birth to the field of nonlinear electrodynamics. Also, such a theory is a low-energy effective action for the electromagnetism of $D$-brane. In fact we can use the effective action of an open string ending on $D$-brane to identify the electromagnetism of $D$-brane \cite{Note1}. One indeed can show that a $n$-dimensional BI action, leads to Monge-gauge-fixed $(p+1)$ dimensional Dirac-Born-Infeld (DBI) action under  dimensional reduction, which for $p=1$ yields the string action \cite{Gibbons:2001gy}. On the other hand BI action is also derivable from that of the open string action \cite{Tseytlin:1999dj}.   
In recent times non-relativistic string theory attracted  considerable attention \cite{Gomis:2020fui}. It is well known that consideration of infinite boost limit in the compactification direction of string theory leads to non-relativistic (NR) behavior \cite{Susskind}. However, this NR theory is a unitary and ultraviolet (UV) complete string theory, containing a Galilean-like global symmetry in flat spacetime. Moreover the spectrum of string excitations satisfy Galilean-invariant dispersion relation. Such a theory is familiar as the NR version of string theory. People found various important theoretical aspects of this stringy limit of string theory (see \cite{Yan}, for a concrete discussion on this direction). For instance, we mention here one of them. The low-energy effective theory corresponding to this NR version yields a Newton-like theory of gravity. Because of its UV completeness, one expects that  one can provide a UV completion of the associated theory of gravity. Therefore NR string can be a successful candidate for a NR theory of quantum gravity and hence may shed light on the ultimate relativistic quantum theory of gravity. Looking at these possibilities, people found interests in the NR version of string theories. Henceforth a huge surge of investigations and efforts emerge to achieve a correct version of NR limit \cite{Yan}. Furthermore we mentioned that BI theory has a close connection with theories of string.  Moreover NR DBI action that arises as the NR limit of string theory describes how strings couple to string Newton–Cartan (SNC) geometry. As a result following the importance of NR strings, one must be interested to analyse the NR limit of BI theory. Therefore understanding the non-relativistic regime of BI theory serves a  two-fold purpose. In one direction we will be able to know about the non-relativistic nature of non-linear electrodynamics and on the other side, various non-relativistic features of open-string can be illuminated. \\ 
\indent Given the importance of the BI theory, we are going to construct a Galilean relativistic version of the leading order BI theory. One can take the limit directly in the full non-linear action. However our purpose is not just
to take the limit but to express everything in non-covariant terms using
electric/magnetic fields and pursue with the analysis. It becomes technically rather complicated. Hence, as a tractable model, the leading
order expansion was treated. The  duality symmetry of the leading order relativistic BI theory has been discussed in \cite{Banerjee03}. We systematically construct the Galilean relativistic action for both electric and magnetic limits, respectively. We derived the equations of motion in terms of Galilean potentials. We then move to the field formulation  and recast the equations of motion in terms of Galilean electric and magnetic fields. We write down a Galilean version of the duality relations. We have found that under these duality relations the equations of motion in the electric limit go to the Bianchi identities of the magnetic limit and the Bianchi identities of the electric limit map to the equations of motion in the magnetic limit. Similarly magnetic limit equations map to that of the electric limit case. We also show the inavariance of the lagrangian under the Galilean boost and gauge transformations. \\
\indent The paper is organised as follows: the mapping relations between Lorentz and Galilean vectors are briefly discussed in section \ref{sec2}. The relativistic BI theory and duality symmetry, in leading order, has been discussed in terms of electric and magnetic fields in section \ref{sec3}. In section \ref{sec4} we give a Galilean covariant action principle for the Born-Infeld theory. The equations of motion in terms of fields are shown in section \ref{sec5}. We discuss the Galilean version of the duality symmetry in section \ref{sec6}. We show that two Galilean limits (electric and magnetic) are swapped under these duality transformations. In section \ref{sec7} the invariance of the Galilean lagrangian under Galilean boost has been shown. The gauge invariance of the lagrangian is shown in section \ref{sec8}. Finally conclusions have been given in section \ref{sec9}. 
\section{Maps relating Lorentz and Galilean vectors} \label{sec2}
\noindent Here we briefly review the basic construction of the scaling relations between special relativistic and Galilean relativistic quantities. We first consider the two Galilean limits (electric and magnetic) for vector quantities.  So first let us take the contravariant vectors. 
Under Lorentz transformations, with boost velocity $u^i$, the time-like and the space-like components of a vector $V^\mu$ transforms as (also considering $u<<c$, so $\gamma \to 1$)
\begin{equation}
    V'^0 = V^0 - \frac{u_j}{c}V^j
    \label{contra1}
\end{equation}
\begin{equation}
    V'^i = V^i - \frac{u^i}{c} V^0
    \label{contra2}
\end{equation}
Galilean vectors $v^0, ~v^i$ are introduced by the scaling,\footnote{Notation: Here relativistic vectors are denoted by capital letters ($V^0, ~V^i$ etc) and Galilean vectors are denoted by lowercase letters
($v^0, ~v^i$ etc)}
\begin{equation}
    V^0 = c v^0, \,\,\,\, V^i = v^i
    \label{contrael}
\end{equation}

\noindent This particular map corresponds to the case $\frac{V^0}{V^i} = c ~\frac{v^0}{v^i}$ in the $c \to \infty$ limit. This yields largely timelike vectors and is called 'electric limit'. 
Now using eqn \ref{contrael} in eqns \ref{contra1} and \ref{contra2} we get
\begin{equation}
v'^0 = v^0
\label{v1}
\end{equation}
\begin{equation}
    v'^i = v^i - u^i v^0
    \label{v2}
\end{equation}
These equations define the usual galilean transformations. \\
\noindent We next consider the magnetic limit which corresponds to largely spacelike vectors
\begin{equation}
    V^0 = -\frac{v^0}{c}, \,\,\,\, V^i = v^i
    \label{contramag}
\end{equation}
Now using \ref{contramag} in \ref{contra1} and \ref{contra2} we get

\begin{equation}
    v'^0 = v^0 + u_j v^j
    \label{v3}
\end{equation}
\begin{equation}
    v'^i = v^i
    \label{v4}
\end{equation}
which is again a Galilean transformation, although the corresponding group is not an invariance group of classical mechanics \cite{Sengupta, Leblond}. Expectedly, the role of time-like and space-like vectors has been reversed from the earlier case. \\ 
\noindent Similarly one can show that the covariant vectors transform under Lorentz transformations in the low velocity limit as
\begin{equation}
    V'_0 = V_0 + \frac{u^i}{c} V_i
    \label{cov1}
\end{equation}
\begin{equation}
    V'_i = V_i + \frac{u_i}{c} V_0
    \label{cov2}
\end{equation}
Now we take the  electric limit in the following way \cite{Bhattacharya}, 
\begin{equation}
    V_0 = \frac{v_0}{c}, \,\,\,\, V_i = v_i
    \label{covel}
\end{equation}
Using \ref{covel} in \ref{cov1} and \ref{cov2} we get
\begin{eqnarray}
  v'_0 = v_0 + u^i v_i
  \label{v5}
\end{eqnarray}
\begin{eqnarray}
  v'_i = v_i
  \label{v6}
\end{eqnarray}
which are again Galilean transformations.\\
\indent Likewise, in the magnetic limit the covariant vectors scale as  
\begin{equation}
    V_0 = -c v_0, \,\,\,\, V_i = v_i
    \label{covmag}
\end{equation}
Using \ref{covmag} in \ref{cov1} and \ref{cov2} we get 
\begin{equation}
    v_0' = v_0
    \label{v7}
\end{equation}
\begin{equation}
    v'_i = v_i - u_i v_0 
    \label{v8}
\end{equation}

\noindent The complete mapping relations are summarised in the table \ref{T1}.
\begin{table}
\caption{Mapping relations}\label{T1}
\begin{center}
\begin{tabular}{|c|c|c|} \hline 
${\rm Limit}$  & $ {\rm Contravariant~~mapping}$ & $ {\rm Covariant~~mapping}$  \\ \hline
${\rm Electric ~~limit}$ & $V^0 \to c~v^0, \,\,  V^i \to v^i$ & $V_0 \to \frac{v_0}{c}, \,\, V_i \to v_i$ \\ \hline
${\rm Magnetic ~~limit}$ & $V^0 \to -\frac{v^0}{c},\,\,V^i \to v^i $ & $V_0 \to -c~v_0. \,\, V_i \to v_i$ \\
\hline
\end{tabular}
\label{T1}
\end{center}
\end{table}

Note that the electric limit of contravariant vectors yields conventional galilean transformations (see (\ref{contrael}) to (\ref{v2})). This is the large time like limit. For the magnetic limit roles of space and time are reversed. Now we have large space like (contravariant) vectors leading to unconventional galilean transformations (\ref{v3}) and  (\ref{v4}).
For covariant components a similar reasoning with appropriate modifications follow. In this case the magnetic limit of covariant vectors  should yield conventional galilean transformations which prompts the scaling (\ref{covel}). Now this is the large time like limit for covariant components. It may be noted that these sets of transformations satisfy the norm preserving condition $V_0V^0+V_iV^i = v_0v^0 + v_iv^i$. Such is very much implicit as the norm for Lorentz vector, given by $V_0V^0+V_iV^i$, will take the form in Galilean limit such that each term will be replaced by the Galilean counter parts $v_0v^0$ and $v_iv^i$.

\section{Born-Infeld theory in the leading order} \label{sec3}
 The Born-Infeld lagrangian is expressed in terms of the square root of a determinant as
\begin{equation}
    \mathcal{L} = -\frac{1}{g^2} \Big(\sqrt{-{\rm det} (\eta_{\mu \nu} + g F_{\mu \nu})} -1 \Big)~,
    \label{BRM1}
\end{equation}
which in the leading order reduces to the following form \cite{Banerjee03}
\begin{equation}
    \mathcal{L} = -\frac{1}{4} F^2 -g^2 (F^2)^2 + 4 g^2 (FFFF)^{\mu}_{~\mu}~.
    \label{molag}
\end{equation}
In the above we denoted $F^2 = F^{\mu \nu} F_{\mu \nu}$ and $(AB)^{\mu}_{~\nu} = A^{\mu \lambda}B_{\lambda \nu}$. 
In principle one can take the NR limit directly in (\ref{BRM1}). However then the analysis becomes technically rather complicated, as our purpose is not just to take the limit but to express everything in noncovariant terms using electric/magnetic fields and pursue with the analysis. It is then advantageous to consider a tractable model. Therefore the leading order expansion of theory is being treated here. 
Using the leading order lagrangian (\ref{molag}) we can write the equations of motion and the Bianchi identities respectively in a covariant form as 
\begin{equation}
    \6_\mu {}^*G^{\mu \nu} = 0, ~~~~~~ \6_\mu {}^*F^{\mu \nu} = 0~, \label{eom1}
\end{equation}
where the dual tensor is defined as,
\begin{equation}
    {}^*G^{\mu \nu} = \frac{1}{2} \epsilon^{\mu \nu \rho \lambda} G_{\rho \lambda}~;
\end{equation}
\begin{equation}
    G^{\mu \nu} = {}^*F^{\mu \nu} + 8g^2 F^2 ({}^*F^{\mu \nu}) + 32g^2 {}^*(FFF)^{\mu \nu}~.
\end{equation}
The set of equations in (\ref{eom1}) is preserved under the following discrete duality transformations
\begin{equation}
    F \to G, ~~~~~~~~~~~~~ G \to -F~.  \label{duality1}
\end{equation}

Since we are eventually going to discuss Galilean relativistic Born-Infeld theory, it is useful to recast the above results in their non-covariant forms using electric and magnetic fields. \\
\noindent \underline{\bf Equations of motion:}\\
\noindent 
The first relation in (\ref{eom1}) yields
\begin{eqnarray}
    \vec \nabla \cdot\vec E + 16 g^2 \vec \nabla \cdot \Big(\vec E (E^2 - B^2)\Big) + 32g^2 \vec B \cdot \vec \nabla (\vec E \cdot \vec B) = 0~; \label{eom2}
    \end{eqnarray}
and    
    \begin{equation}
        \begin{split}
            -\6_t \vec E + (\vec \nabla \times \vec B) - 16g^2 \6_t \Big(\vec E (E^2 - B^2)  \Big)  -32g^2 \6_t \Big(\vec B (\vec E \cdot \vec B)\Big) &\\  - 16g^2 \vec \nabla \times \Big(\vec B (E^2 - B^2)  \Big)  -32g^2 \vec \nabla \times \Big(\vec E \times (\vec E \times \vec B)  \Big) -32 g^2 \vec \nabla \times (B^2 \vec B) = 0~. \label{eom3}
        \end{split}
    \end{equation}
Here the electric and magnetic fields are defined as,
    \begin{eqnarray}
        E^i = F^{0i} = \6^0 A^i - \6^i A^0, \nonumber\\
        B^i = \frac{1}{2} \epsilon^{ijk} F_{jk} = \epsilon^{ijk} \6_j A_k~. \label{EBD}
    \end{eqnarray}
  Note that eqns \ref{eom2} and \ref{eom3} incorporate nonlinear terms apart from the standard Maxwell ones. 
  
\noindent \underline{\bf Bianchi identities:}\\
The other relation in (\ref{eom1}) leads to two of the standard Maxwell's four equations:
\begin{align}
     \vec \nabla \times \vec E  =  -\frac{\6 \vec B}{\6 t}~; \label{bianchi1} \\
     \vec \nabla \cdot \vec B = 0~. \label{bianchi2}
\end{align}
Expectedly, the Bianchi identities retain their usual Maxwell form. The equations of motion in the limit $g \to 0$ also reproduce the desired Maxwell equations. 
The duality relations (\ref{duality1}) can be expressed explicitly in the component form as well. \\
\noindent \underline{\bf Duality relations:}\\
The first one in (\ref{duality1}) implies the following two mappings:
\begin{eqnarray}
    &&\vec E \to \vec B - 16 g^2 (E^2 - B^2) \vec B - 32 g^2 \vec E \times (\vec E \times \vec B) -32g^2 B^2 \vec B~; \label{d1}\\
    &&\vec B \to -\vec E - 16 g^2(E^2 - B^2) \vec E - 32g^2 (\vec E \cdot \vec B)\vec B~. \label{d2}
\end{eqnarray}
Whereas the last one signifies
\begin{eqnarray}
    &&\vec B - 16 g^2 (E^2 - B^2) \vec B - 32 g^2 \vec E \times (\vec E \times \vec B) -32g^2 B^2 \vec B \to - \vec E~; \label{d3} \\
    &&\vec E + 16 g^2(E^2 - B^2) \vec E + 32g^2 (\vec E \cdot \vec B)\vec B \to \vec B~. \label{d4}
\end{eqnarray}

It can be shown that the two sets of dualities given above i.e. (\ref{d1}), (\ref{d2}) and (\ref{d3}), (\ref{d4}) are consistent among each other.  Also in the limit $g \to 0$ corresponding to Maxwell theory, the results yield the well known duality maps,
\begin{equation}
    \vec E \to \vec B, ~~~~~~~~~~~ \vec B \to -\vec E
\end{equation}
using either set.  Under these duality relations the equations of motion i.e. (\ref{eom2}) and (\ref{eom3}) will go to the Bianchi identities (\ref{bianchi2}) and (\ref{bianchi1}) respectively, and vice-versa.

\section{Galilean relativistic Born-Infeld theory: action principle} \label{sec4}
 Here we wish to systematically build up a Galilean relativistic version of the Born-Infeld theory. Let us now start from the relativistic Born-Infeld theory described by the Lagrangian given in (\ref{molag}). Below we consider two Galilean limits separately. 
\subsection{Electric limit}
The calculation will be done following a particular general idea. Before taking $c \to \infty$ limit, we will write down all the terms (if that term has any possible $c$ dependence) in any mathematical expression in terms of $c$. So relativistic four vectors can be written in terms of Galilean vectors with possible $c$ dependence using Table \ref{T1} and the partial derivatives i.e. $\partial^0, ~\partial^i, ~\partial_0, ~\partial_i$ will follow the relation: $\partial^0 = - \partial_0 = -\frac{1}{c} \partial_t$ and $\partial^i = \partial_i$. These are dictated by the relations $x^0=-x_0=ct, x^i=x_i$. Finally the $c\to\infty$ limit will be considered. 
Then using the relations given in Table \ref{T1} we can express the terms in (\ref{molag}) as
\begin{eqnarray}
F^2 
\xrightarrow[\text{$c \to \infty$}]{\text{Electric limit}}  -2 \6^i a^0 \Big( \partial_t a_i - \partial_i a_0  \big) + f^2, \hspace{1in}
\label{part1}
\end{eqnarray}
and
\begin{eqnarray}
    (FFFF)^{\mu}_{~\mu}  \xrightarrow[\text{$c \to \infty$}]{\text{Electric limit}} 2 \6^l a^0 \6^k a^0 (\6_l a_0 - \6_t a_l)(\6_k a_0 - \6_t a_k) + 4 \6^i a^0 (\6_t a_m - \6_m a_0) (ff)^m_{~i}\nonumber \\+ (ffff)^i_{~i}~;
\end{eqnarray}
where we denoted $f^2 = f^{ij}f_{ij}$.
So in the electric limit we can write down the Galilean relativistic action as
\begin{equation}
    S_e = \int \mathcal{L}_e dt d^3 x~,
    \label{se}
\end{equation}
where the Lagrangian takes the following form
\begin{eqnarray}
    \mathcal{L}_e = \frac{1}{2} \6^i a^0 \Big( \partial_t a_i - \partial_i a_0  \big) -\frac{1}{4} f^2 -g^2 \Big(-2 \6^i a^0 \Big( \partial_t a_i - \partial_i a_0  \big) + f^2 \Big)^2 + \nonumber \hspace{0.8in}\\ 4g^2 \Big( 2 \6^l a^0 \6^k a^0 (\6_l a_0 - \6_t a_l)(\6_k a_0 - \6_t a_k) + 4 \6^i a^0 (\6_t a_m - \6_m a_0) (ff)^m_{~i} + (ffff)^i_{~i}   \Big)~. \label{le}
\end{eqnarray}
Now varying the action (\ref{se}) with respect to $a_0, ~a_j, ~a^0, ~a^j$ we get the following equations respectively,
\begin{eqnarray}
    \6_i \6^i a^0 + 8g^2 \6_i (\alpha \6^i a^0) - 32 g^2 \6_i\Big(\6^i a^0 \6^k a^0 (\6_k a_0 - \6_t a_k) - \6^j a^0 (ff)^i_{~j}   \Big) = 0~; 
    \label{ee1}
\end{eqnarray}
\begin{eqnarray}
    \6_t \6^j a^0 - \6_i f^{ij} + 8 g^2 \6_t (\alpha \6^j a^0) - 8 g^2 \6_i (\alpha f^{ij}) + 32 g^2 \6_t \Big( \6^j a^0 \6^l a^0 (\6_l a_0 - \6_t a_l) + \6^l a^0 (ff)_l^{~j} \Big) + \nonumber \\ 8g^2 \6_i \Big(4 \6^i a^0 (\6_t a_k - \6_k a_0) f^{jk} - 4 \6^j a^0 (\6_t a_k - \6_k a_0)  f^{ik} + 2 (fff)^{ji} - 2 (fff)^{ij}  \Big) = 0~; 
    \label{ee2}
\end{eqnarray}
\begin{eqnarray}
    \6^i (\6_t a_i - \6_i a_0) + 8g^2 \6^i \Big(\alpha (\6_t a_i - \6_i a_0)  \Big) + 32 g^2 \6^i \Big( \6^j a^0 (\6_i a_0 - \6_t a_i) (\6_j a_0 - \6_t a_j) + (\6_t a_k - \6_k a_0) (ff)_i^{~k}  \Big) = 0~;
    \label{ee3}
\end{eqnarray}
and
\begin{eqnarray}
    \6^i f_{ij} + 8g^2 \6^i (\alpha f_{ij}) - 8g^2 \6^i \Big( 4\6^k a^0 (\6_t a_i - \6_i a_0) f_{jk} - 4 \6^k a^0 (\6_t a_j - \6_j a_0) f_{ik} + 2 (fff)_{ji} - 2 (fff)_{ij}   \Big) = 0~. 
    \label{ee4}
\end{eqnarray}
In the above $\alpha$ represents 
\begin{equation}
    \alpha  = -2 \6^i a^0 \Big( \partial_t a_i - \partial_i a_0  \big) + f^2~.
\end{equation}

\subsection{Magnetic limit}
 In the magnetic limit, using Table \ref{T1}, we can write down the Galilean relativistic action as
\begin{equation}
    S_m = \int \mathcal{L}_m dt d^3 x~,
    \label{sm}
\end{equation}
where the Lagrangian takes the following form
\begin{eqnarray}
    \mathcal{L}_m = \frac{1}{2} \6_i a_0 (\6_t a^i - \6^i a^0) -\frac{1}{4} f^2 - g^2 \Big( -2 \6_i a_0 (\6_t a^i - \6^i a^0) + f^2 \Big) + \nonumber \hspace{0.5in} \\ 4g^2 \Big(2 \6_i a_0 \6_j a_0 (\6_t a^i - \6^i a_0) (\6_t a^j - \6^j a^0) + 4 \6_k a_0 (\6_t a^i - \6^i a^0) (ff)_i^{~k} + (ffff)^i_{~i}  \Big)~. \label{lm}
\end{eqnarray}
Now varying the action (\ref{sm}) with respect to $a_0, ~a_j, ~a^0, ~a^j$ we have the following set of equations respectively, 
\begin{eqnarray}
    \6_i (\6_t a^i - \6^i a^0) + 8g^2 \6_i \Big(\alpha  (\6_t a^i - \6^i a^0) \Big) + 32g^2 \6_i \Big( \6_j a_0 (\6_t a^i - \6^i a^0) (\6_t a^j - \6^j a^0) + (\6_t a^l - \6^l a^0) (ff)_l^{~i}  \Big) = 0~;
    \label{em1}
\end{eqnarray}
\begin{eqnarray}
    \6_i f^{ij} + 8g^2 \6_i (\alpha f^{ij}) -8g^2 \6_i \Big( 4 \6_k a_0 (\6_t a^i - \6^i a^0) f^{jk} - 4 \6_k a_0 (\6_t a^j - \6^j a^0) f^{ik} + 2 (fff)^{ji} - 2 (fff)^{ij} \Big) = 0~; 
    \label{em2}
\end{eqnarray}
\begin{eqnarray}
    \6^i \6_i a_0 + 8g^2 \6^i (\alpha \6_i a_0) -32g^2 \6^i \Big(\6_i a_0 \6_k a_0 (\6^k a^0 - \6_t a^k) - \6_j a_0 (ff)^i_{~j}   \Big) = 0~; 
    \label{em3}
\end{eqnarray}
and
\begin{eqnarray}
    \6_t \6_j a_0 - \6^i f_{ij} + 8g^2 \6_t (\alpha \6_j a_0) - 8 g^2 \6^i (\alpha f_{ij}) + 32g^2 \6_t \Big(\6_j a_0 \6_i a_0 (\6_t a^i - \6^i a^0) + \6_k a_0 f_{jm} f^{mk}   \Big) + \nonumber \\ 8 g^2 \6^i \Big( 4 \6_i a_0 (\6_t a^k - \6^k a^0) f_{jk} - 4\6_j a_0 (\6_t a^k - \6^k a^0) f_{ik}  + 2 (fff)_{ji} - 2 (fff)_{ij} \Big) = 0~. 
    \label{em4}
\end{eqnarray}

We observe that the electric limit equations of motion ((\ref{ee1}) - (\ref{ee4})) and magnetic limit equations of motion ((\ref{em1}) - (\ref{em4})) satisfy a symmetry property. Under an interchange of covariant and contravariant indices, the results for electric limit go to magnetic limit and vice versa. 

\hspace{0.8in}
\fbox{\begin{minipage}{30em}
\begin{equation*}
    {\rm {\bf  Contravariant \longleftrightarrow Covariant}} \Longleftrightarrow {\rm {\bf Electric \longleftrightarrow Magnetic}}
\end{equation*}
\end{minipage}}
\vspace{0.5in}
\section{Galilean electric and magnetic fields} \label{sec5}
\noindent Here we introduce the Galilean limit of electric and magnetic fields and write down the equations of motion for Galilean relativistic Born-Infeld theory, obtained in the previous section, in terms of these fields. Note that relativistic electric and magnetic fields are defined in (\ref{EBD}).  

\subsection{Electric limit}
\noindent Using the maps given in table \ref{T1} we can write the contravariant and covariant components of the electric field as 
\begin{eqnarray}
E^i = -\frac{1}{c} \6_t a^i - c \6^i a^0, ~~~~ E_i = \Big(\frac{1}{c} \6_t a_i - \frac{1}{c} \6_i   a_0\Big)~. 
\end{eqnarray}
Also we define the Galilean electric and magnetic fields as 
\begin{equation}
    e^i = \lim_{c \to \infty} \frac{E^i}{c} = -\6^i a^0, \,\,\,~ b^i =  \lim_{c \to \infty} B^i = \epsilon^{ij}_{~k} \6_j a^k, ~~~~e_i = \lim_{c \to \infty} c E_i = (\6_t a_i - \6_i a_0), \,\,\,~~ b_i = \lim_{c \to \infty} B_i = \epsilon_{i}^{~jk} \6_j a_k~.  \label{emapping1}
\end{equation}
Note that the contravariant component $e^i$ in the electric limit represents the electric field as given by Bellac and Levy-Leblond. On the other hand, the covariant component $e_i$ in the magnetic limit also represents the electric field. They have similar gauge transformation properties. Here in this paper we have given a systematic action formalism for Galilean relativistic Born-Infeld theory. To construct the Galilean relativistic action we need both covariant and contravariant vectors since these are distinct entities in the Galilean theory, not being connected by any metric. If we replace the covariant components by contravariant ones in the electric limit case we will end up with the magnetic limit case and vice-versa. This fact manifests itself only if we consider the covariant and contravariant sectors separately as we have done here.The interplay between the covariant and the contravaraint indices that leads to an interchange of the electric and the magnetic limits of the theory is a new feature observed first in the Galilean Maxwell case in reference \cite{Bhattacharya} as well as here. Now regarding the physical meaning of covariant electric field $e_i$ (or magnetic field $b_i$) one can say that the covariant electric field (or magnetic field) in the electric limit behaves identically to contravariant electric field (or magnetic field) in the magnetic limit and vice-versa.
Now we write the field equations that we derived in the previous section  in terms of the Galilean electric and magnetic fields. Using the maps given in Eq.  (\ref{emapping1}) we express (\ref{ee1}) as 
\begin{equation}
    \6_i e^i + 16g^2 \6_i\Big( (e^2 - b^2)  e^i \Big) + 32g^2  b^i\6_i(e^j b_j) = 0~, \label{efe1}
\end{equation}
where $e^2 = e^i e_i$, $b^2 = b^i b_i$.
Similarly (\ref{ee2}) takes the form as follows: 
\begin{eqnarray}
    -\6_t  e^i + (\vec \nabla \times \vec b)^i - 16 g^2 \6_t \Big((e^2 - b^2) e^i \Big) -32 g^2 \6_t\Big( b^i( e^k b_k) \Big) - 16g^2 [\vec \nabla \times \Big((e^2 - b^2)\vec b  \Big)]^i \nonumber \\ -32g^2 [\vec \nabla \times \Big(\vec e \times (\vec e \times \vec b) \Big)]^i -32g^2 [\vec \nabla \times (b^2 \vec b)]^i = 0~. \label{efe2}
\end{eqnarray}
In an identical manner we can recast equations (\ref{ee3}) and (\ref{ee4}) in terms of Galilean covariant electric and magnetic fields, defined in Eq.  (\ref{emapping1}). We are not presenting them here as one can can easily check that they are of same form as those in magnetic limit, with the covariant fields replaced by contravariant fields (as defined in (\ref{bmapping2})). This we leave for the next subsection. This fact verifies our observation which is mentioned at the end of Sect. \ref{sec4}, that under the interchange of covariant and contravariant indices electric limit goes to magnetic limit and vice-versa.  

We next write the Bianchi identities using fields. We can see from relativistic Born-Infeld theory that Bianchi identities ((\ref{bianchi1}) and (\ref{bianchi2})) retain the same form as that of Maxwell case. So in the Galilean limit these Bianchi identities reduce to those of Galilean relativistic Maxwell theory, shown in \cite{Bhattacharya}.  
First, using the field definitions given in (\ref{emapping1}) we can write 
\begin{equation}
\vec \nabla \times \vec e = \epsilon^{ij}_{~~k}\6_j e^k = \epsilon^{ij}_{~~k}\6_j \6^k a^0 = 0~. \label{ebe1}
\end{equation}
The same result follows by starting from (\ref{bianchi1}), after an appropriate insertion of $ c$ (the velocity of light in free space),
\begin{equation}
    \vec \nabla \times \vec E = - \frac{1}{c} \frac{\6 \vec B}{ \6 t}
\end{equation}
and then using (\ref{emapping1}) followed by $c \to \infty$,
\begin{equation}
    \vec \nabla \times \vec e = - \frac{1}{c^2} \frac{\6 \vec b}{\6 t} \to 0~.
\end{equation}
Similarly one finds,
\begin{equation}
    \vec\nabla \cdot \vec b = \6_i b^i =  0~. \label{ebe2}
\end{equation}

On the other hand if we start with the relativistic equations directly, then using the mapping relation summarised in table \ref{T05} in (\ref{eom2}) we have
\begin{eqnarray}
   c ~\6_i e^i + 16g^2 c ~\6_i( e^i (e^2 - b^2)  \Big) + 32g^2 c ~b^i \6_i( e^k b_k) = 0~,
\end{eqnarray}
which in the $c \to \infty$ will give rise to  (\ref{efe1}). Similarly from (\ref{eom3}) we can reproduce (\ref{efe2}) in the $c \to \infty$ limit. This shows the internal consistency of the results.

\begin{table}
\caption{Fields in galilean limit}\label{T05}
\begin{center}
\begin{tabular}{|c|c|c|} \hline 
${\rm Limits}$  & $ {\rm Electric ~field}$ & $ {\rm Magnetic ~field}$  \\ \hline
${\rm Electric ~limit}$ & $E^i \to c e^i, ~E_i \to \frac{e_i}{c}$ & $B^i \to b^i, ~B_i \to b_i$ \\ \hline
${\rm Magnetic ~limit }$ &  $ E^i \to \frac{e^i}{c}, ~E_i \to c e_i$  & $B^i \to b^i, ~B_i \to b_i $ \\
\hline
\end{tabular}
\label{T05}
\end{center}
\end{table}
\subsection{Magnetic limit}
The contravariant and covariant components of the electric field can be written in this limit from the mapping relations in table \ref{T1} as 
\begin{equation}
    E^i = -\frac{1}{c} \6_t a^i + \frac{1}{c} \6^i a^0, ~~~~~~~ E_i = \Big(\frac{1}{c} \6_t a_i - c \6_i a_0 \Big)~.
\end{equation}
Correspondingly, the Galilean electric and magnetic fields are defined as, 
\begin{equation}
   e^i = \lim_{c \to \infty} c E^i = -(\6_t a^i - \6^i a^0), \,\,\, ~~ b^i = \lim_{c \to \infty} B^i = \epsilon^{ij}_{~k} \6_j a^k ~~~~~~
  e_i = \lim_{c \to \infty} \frac{E_i}{c} = -\6_i a_0, \,\,\, ~~b_i = \lim_{c \to \infty} B_i = \epsilon_{i}^{~jk} \6_j a_k~.  \label{bmapping2}
\end{equation}
Now we write the field equations that we derived earlier in terms of the Galilean electric and magnetic fields. From equation (\ref{em1}) using the mapping relation given in (\ref{bmapping2}) we have
\begin{equation}
    \6_i e^i + 16g^2 \6_i \Big( e^i (e^2 - b^2)  \Big) + 32g^2  b^i \6_i( e^k b_k) = 0~. \label{mfe1}
\end{equation}
Similarly (\ref{em2}) yields
\begin{equation}
    (\vec \nabla \times \vec b) + 16g^2 \vec \nabla \times \Big((e^2 - b^2) \vec b  \Big) - 32g^2 \vec \nabla \times \Big( \vec e (\vec e \cdot \vec b) \Big) = 0~. \label{mfe2}
\end{equation}
From the Bianchi identities we have the following two equations
\begin{equation}
    (\vec \nabla \times \vec e)^i = \epsilon^{ij}_{~~k}\6_j e^k =  -\6_t  b^i~; \label{mbe1}
\end{equation}
and
\begin{equation}
    \vec \nabla \cdot \vec b = \6_i b^i = 0~. \label{mbe2}
\end{equation}
These equations can also be derived directly from the relativistic Maxwell equations, as we did for the electric case. 

\section{Duality transformations} \label{sec6}
The duality transformations given in (\ref{d1}) and (\ref{d2}) will take the following forms in the non-relativistic limit
\begin{eqnarray}
     e^i \to  b^i + 16 g^2 (e^2 - b^2)  b^i - 32 g^2  e^i ( e_k b^k) \label{nd1} \\
     b^i \to - e^i - 16 g^2 (e^2 - b^2)  e^i - 32g^2 ( e^k b_k) b^i~.
     \label{nd2}
\end{eqnarray}
The forms are different in the two limits although at first glance it might not appear. Note that these equations contain expressions like $e_k b^k$ and $e^k b_k$ which are distinct
entities in the two limits.
Using them in the electric and magnetic limits switches the expressions from one to the other.
Now we use these duality relations in the electric and magnetic limit equations, obtained in the last section, separately and observe the consequences. 

Let us start with the equations in electric limit.
Using the non-relativistic duality relations (\ref{nd1}) and (\ref{nd2}) in (\ref{efe1}) and keeping upto $O(g^2)$ we get
\begin{equation}
    \vec \nabla\cdot\vec b = \6_i b^i = 0~. \label{du1}
\end{equation}
Similarly use of them in (\ref{efe2}) gives rise to
\begin{equation}
    (\vec \nabla \times \vec e)^i = \epsilon^{ij}_k \6_j e^k = - \6_t b^i~. \label{du2}
\end{equation}
Whereas under these duality transformations the first Bianchi identity (\ref{ebe1}) leads to
\begin{equation}
   (\vec \nabla \times \vec b) + 16g^2 \vec \nabla \times \Big((e^2 - b^2) \vec b  \Big) - 32g^2 \vec \nabla \times \Big( \vec e (\vec e \cdot \vec b) \Big) = 0~,  \label{du3}
\end{equation}
while the second Bianchi identity (\ref{ebe2}) yields
\begin{equation}
   \6_i e^i + 16g^2 \6_i( e^i (e^2 - b^2)  \Big) + 32g^2  b^i \6_i( e^k b_k) = 0~. \label{du4}
\end{equation}
Interestingly, these relations ((\ref{du1}) - (\ref{du4})) exactly correspond to those in magnetic limit (see, (\ref{mfe1}) - (\ref{mbe2})). 

Now we will concentrate on the equations in magnetic limit.
Using the non-relativistic duality relations (\ref{nd1}) and (\ref{nd2}) in (\ref{mfe1}) and keeping upto $O(g^2)$ we get
\begin{equation}
    \vec \nabla \cdot \vec b = \6_i b^i = 0~. \label{du5}
\end{equation}
Again using of the duality relations in  (\ref{mfe2}) will give rise to
\begin{equation}
    (\vec \nabla \times \vec e)^i = \epsilon^{ij}_k \6_j e^k = 0~.
    \label{du6}
\end{equation}
Under these the first Bianchi identity (\ref{mbe1}) reduces to
\begin{eqnarray}
    -\6_t \vec e + (\vec \nabla \times \vec b) - 16 g^2 \6_t \Big((e^2 - b^2)\vec e \Big) -32 g^2 \6_t\Big(\vec b(\vec e \cdot \vec b) \Big) - 16g^2 \vec \nabla \times \Big((e^2 - b^2)\vec b  \Big) \nonumber \\ -32g^2 \vec \nabla \times \Big(\vec e \times (\vec e \times \vec b) \Big) -32g^2 \vec \nabla \times (b^2 \vec b) = 0~,
    \label{du7}
\end{eqnarray}
while the second Bianchi identity (\ref{mbe2}) leads to
\begin{equation}
     \6_i e^i + 16g^2 \6_i\Big( (e^2 - b^2)  e^i \Big) + 32g^2  b^i\6_i(e^j b_j) = 0~.  \label{du8}
\end{equation}
Here we see that these relations ((\ref{du5})-(\ref{du8})) exactly correspond to the results (\ref{efe1}-\ref{ebe2}) found for the electric limit.

This shows that the results for {\it the electric and magnetic limits are swapped under the non-relativistic duality transformations (\ref{nd1}) and (\ref{nd2})}.

\section{Invariance under Galilean boost transformations} \label{sec7}
 Here we discuss the invariance of the Galilean Lagrangians under the Galilean boost transformation. We will consider both electric and magnetic limit cases separately.
 
Using the electric limit scaling given in table \ref{T1}, we find that the Galilean electric and magnetic fields transform under the Galilean boost as follows 
\begin{eqnarray}
 e'^i = e^i,  ~~~b'^i =  b^i -\Big(\vec v \times \vec e \Big)^i, ~~~~e'_i = e_i + \Big(\vec v \times \vec b  \Big)_i, ~~~ b'_i = b_i~.
\end{eqnarray}
Now considering the variation we get
\begin{equation}
    \delta e^i = 0, ~~\delta b^i = -\Big(\vec v \times \vec e \Big)^i, ~~~~~ \delta e_i =  \Big(\vec v \times \vec b  \Big)_i, ~~\delta b_i = 0~.
    \label{vf}
\end{equation}
We can write down the Lagrangian in terms of fields as
\begin{equation}
    \mathcal{L}_e = \frac{1}{2} (e^i e_i - b^i b_i) + 4g^2 (e^i e_i - b^i b_i)^2 + 4 g^2 \Big( 2 (e^i e_i)(e^j e_j) -4 (e^i e_i) (b^j b_j) + 4 (e^i b_i) (e_j b^j) + 2 (b^i b_i) (b^j b_j)  \Big)~.
    \label{lfe}
\end{equation}
Now computation of the variation of this Lagrangian by using (\ref{vf}) one can easily verify that 
\begin{equation}
    \delta \mathcal{L}_e = 0~. 
    \label{dlfe}
\end{equation}

In a similar manner using the magnetic limit scaling given in table \ref{T1}, we find that the 
\begin{equation}
    \delta \mathcal{L}_m = 0~. 
\end{equation}

\section{Gauge invariance} \label{sec8}
\noindent It is well known that the relativistic Born-Infeld Lagrangian \ref{molag} is invariant under the following gauge transformation,
\begin{equation}
    \delta A_{\mu} = \6_{\mu} \alpha, \,\,\,  \delta A^{\mu} = \6^{\mu} \alpha ~,
    \label{cpot}
\end{equation}
where $\alpha$ is an arbitrary function of spacetime.
Below we will explore how the same is maintained at Galilean level.
Here we can consider a relatively more general gauge transformation of the following form:
\begin{equation}
    \delta A_{\mu} = \6_{\mu} \alpha, \,\,\,  \delta A^{\mu} = \6^{\mu} \beta~. \label{cpot1}
\end{equation}
Note that in the relativistic theory the covaraint and contravariant vectors are related by a metric implying $\alpha = \beta$. However, this is not true in Galilean limit due to the lack of a non-degenerate metric. Hence we are free to choose $\alpha \neq \beta$. We now discuss the electric and magnetic limit cases separately.

\noindent \underline {\bf Electric limit} \\
From (\ref{cpot1}) and using the mapping relations given in table \ref{T1}  we deduce the following relations,
\begin{eqnarray}
\delta A_0 = \6_0 \alpha 
\implies \delta a_0 = \6_t \alpha~; \label{eg1}
\end{eqnarray}
\begin{eqnarray}
\delta A_i = \6_i \alpha \implies \delta a_i = \6_i \alpha~; \label{eg2}
\end{eqnarray}
\begin{eqnarray}
\delta A^0 = \6^0 \beta \implies c \delta a^0 = -\frac{1}{c} \6_t \beta   
\xrightarrow[\text{$c \to \infty$}]{\text{}} \delta a^0 = 0~; 
\label{eg3}
\end{eqnarray}
and
\begin{eqnarray}
\delta A^i = \6^i \beta \implies \delta a^i = \6^i \beta~.
\label{eg4}
\end{eqnarray}
These results are summarised in table \ref{T3}. Varying the Lagrangian and using the results summarised in table \ref{T3} we can show that
\begin{equation}
    \delta \mathcal{L}_e = 0~. 
\end{equation}
\begin{table}
\caption{Variations of the Galilean potentials}\label{T3}
\begin{center}
\begin{tabular}{|c|c|c|} \hline 
${\rm Variable}$  & $ {\rm Electric ~~limit}$ & $ {\rm Magnetic ~~limit}$  \\ \hline
$a^0$ & $\delta a^0 = 0$ & $\delta a^0 = \6_t \beta$\\ \hline
$a^i$ & $\delta a^i = \6^i \beta$  & $\delta a^i = \6^i \beta$ \\ \hline
$a_0$ & $\delta a_0 = \6_t \alpha$ & $\delta a_0 = 0 $\\ \hline
$a_i$ & $\delta a_i = \6_i \alpha$  & $\delta a_i = \6_i \alpha$\\ 
\hline
\end{tabular}
\label{T3}
\end{center}
\end{table}

\noindent \underline {\bf Magnetic limit} \\
\noindent Proceeding in a similar way as in the electric limit we can derive the variations of the Galilean potentials in the magnetic limit which are summarised in table \ref{T3}. Using these results and considering the variation of the Lagrangian \ref{lm} we can show that 
\begin{equation}
    \delta \mathcal{L}_m = 0~.
\end{equation}
This shows that the Galilean version of BI theory, both in electric and magnetic limits, is gauge invariant. Observe that various limits are related as: \\

\fbox{\begin{minipage}{35em}
\begin{equation*}
    {\rm {\bf  Contravariant \longleftrightarrow Covariant}} \Longleftrightarrow {\rm {\bf Electric \longleftrightarrow Magnetic}, ~~{\bf \alpha} \longleftrightarrow {\bf \beta}}
\end{equation*}
\end{minipage}}

\section{Conclusions} \label{sec9}
\noindent Galilean relativistic electrodynamics always play an important role to describe electrodynamics of moving bodies with low velocities. This construction helps us to understand a wide range of physics phenomena that we experience in our day to day life \cite{Germain}. So in that sense it is a more practical approach than relativistic electrodynamics which is not so well adapted and less efficient than the
non-relativistic limits. While such limits have a domain of restricted
validity, they are effective in the practical explanation of Galilean phenomena \cite{Germain}. 
Electrodynamics of continuous media at low velocities provides a nice example where the efficacy of the two non-relativistic limits comes to the fore. In reference [27] two approximate Galilean sets of Maxwell equations in continuous media were postulated leading to the field transformations given in the table \ref{T2} for a velocity small compared to light. These relations, for both limits, agree with the expressions derived in reference [18] \footnote{See the equations given in table V of reference [18]} using our approach.
    \begin{table}
\caption{Field transformations}\label{T2}
\begin{center}
\begin{tabular}{|c|c|} \hline 
${\rm Electric ~~limit}$ & ${\rm Magnetic ~~limit}$ \\ \hline
$\rho_e = \rho'_e$ & $\rho_m = \rho'_m + \frac{\vec v \cdot \vec j'_m}{c^2}$ \\ \hline
$\vec j_e = \vec j'_e - \vec v \rho'_e$ & $\vec j_m = \vec j'_m$ \\ \hline
$\vec B_e = \vec B'_e + \frac{\vec v \times \vec E'_e}{c^2}$ & $\vec B_m = \vec B'_m$ \\ \hline
$\vec E_e = \vec E'_e$ & $\vec E_m = \vec E'_m - (\vec v \times \vec B'_m)$ \\ \hline
\end{tabular}
\label{T2}
\end{center}
\end{table} In the same vein the study of Galilean relativistic nonlinear electrodynamics is also useful. This is further enhanced by its connection with Galilean relativistic string theory. \\
\indent A systematic construction of the Galilean relativistic Born-Infeld (BI) theory using an action principle has been provided here. BI theory which is a non-linear theory of electromagnetism, plays an important role in the relativistic string theory as the electrodynamics on the D-brane is explained by the BI theory. The Galilean relativistic BI theory emerges in the studies of non-relativistic open strings. There is no systematic study of Galilean relativistic BI theory available in the literature. Here in this paper we have studied quite extensively the Galilean relativistic BI theory in the leading order and various symmetries involved with it.\\
\indent We know that BI theory satisfies a duality symmetry. We have considered the leading order expressions. The conventional covariant expressions have been expressed in component forms using electric and magnetic fields. The duality symmetry was discussed in terms of these fields. We then discuss two Galilean limits (electric and magnetic) separately and write down the equations of motion in terms of potentials. We consider both contravariant and covariant potentials as they represent different quantities in the Galilean limit. We observe that under the interchange of covariant and contravariant indices the equations of motion in the electric limit go to that of magnetic limit and vice-versa. We then recast these equations of motion in terms of Galilean electric and magnetic fields. This is necessary for discussing the electric-magnetic duality. We express the duality relations in terms of Galilean electric and magnetic fields up to leading order. We show explicitly that under these duality transformations the electric and magnetic limit equations are swapped; i.e. electric limit equations of motion go to magnetic limit Bianchi identities and electric limit Bianchi identities go to magnetic limit equations of motion and vice-versa. This is an important feature and can be used to find new solutions in the magnetic limit from known solutions in the electric limit and vice-versa. Next we show that both electric and magnetic limit Lagrangians are invariant under Galilean boost transformations. In the end we show that these Galilean relativistic Lagrangians are gauge invariant. We believe our results are important and will shed some light on various aspects of non-relativistic string theory.\\
\indent There have been studies on nonlinear electromagnetism defined on non-commutative spaces \cite{Banerjee03}. It will be quite interesting to study the Galilean relativistic aspects of nonlinear electromagnetism defined on non-commutative spaces using the method developed here. It is also interesting to study the effects of this electric-magnetic duality symmetry in non-relativistic string theory. We hope we will address these issues in near future.

\section{Acknowledgements}
\noindent Two of the authors (RB and SB) acknowledge the support from a DAE Raja Ramanna Fellowship (grant no: \\$1003/(6)/2021/RRF/R\&D-II/4031$, dated: $20/03/2021$). The other author (BRM) is supported by  Science and Engineering Research Board (SERB), Department of Science $\&$ Technology (DST), Government of India, under the scheme Core Research Grant (File no. CRG/2020/000616). 

\end{document}